\documentclass[10pt,letterpaper]{article}

\usepackage{opex3}
\usepackage{cite}

\begin{document}

\title{Optical lock-in particle tracking in optical tweezers}  
\author{Michael~A~Taylor, Joachim~Knittel and Warwick~P~Bowen$^{*}$}
\address{Centre for Engineered Quantum Systems, University of Queensland, St Lucia, Queensland 4072, Australia }
\email{$^*$ wbowen@physics.uq.edu.au}   

\date{\today}


\begin{abstract}
 We demonstrate a lock-in particle tracking scheme in optical tweezers based on stroboscopic modulation of an illuminating optical field. This scheme is found to evade low frequency noise sources while otherwise producing an equivalent position measurement to continuous measurement. This was demonstrated, and found to yield on average 20~dB of noise suppression in the frequency range 10--5000~Hz, where low frequency laser noise and electronic noise was significant, and 35~dB of noise suppression in the range 550--710~kHz where laser relaxation oscillations introduced laser noise. The setup is simple, and compatible with any trapping optics.
\end{abstract}


\section{Introduction}

 In optical tweezers, small particles are trapped by the electric field gradient at the focus of a tightly focused laser beam~\cite{Ashkin1987}. After the light interacts with the particle, it encodes information about the particle position~\cite{Taylor2012QNL}, which is typically extracted with a quadrant detector at the back-focal plane of a condenser lens~\cite{Gittes1998}. This allows particle tracking with sub-nanometer sensitivity~\cite{Chavez2008} as forces ranging from subpiconewton~\cite{Moffitt2008,Finer1994} to nanonewton~\cite{Jannasch2012} are controllably applied. Such experiments have uncovered important phenomena in biophysics,  including both the dynamics and magnitude of the forces applied by biological motors~\cite{Finer1994,Svoboda1993,Bustamante2004},  the stretching and folding properties of DNA and RNA~\cite{Bustamante2004,Bustamante2005,Greenleaf2006}, the dynamics of virus-host coupling~\cite{Kukura2009}, the strain on an enzyme during catalysis~\cite{Bustamante2004}, and the rheological properties of cellular cytoplasm~\cite{Taylor2012_squeezing,Yamada2000,Senning2010,Norrelykke2004}. Furthermore, optical tweezers provide a significant tool for studying the fundamental physics of Brownian motion~\cite{Franosch2011,Huang2011} and quantum optomechanics~\cite{Chang2010,Li2011}.

  While shot-noise establishes the fundamental sensitivity limit for optical tweezers based measurements~\cite{Taylor2012QNL,Taylor2011,Tay2009}, real experiments are generally limited by technical noise sources such as laser noise, electronic noise in the detector, or drifts of mirrors in the experiment. These technical sources of error can be a significant hindrance to precision measurement, so much effort has gone into reducing them~\cite{Chavez2008,Moffitt2008,Kukura2009,Taylor2011}. Recently, an optical lock-in particle tracking scheme was developed which allowed evasion of low-frequency technical noise without needing to remove the noise sources from the experiment. This was applied in conjunction with quantum correlated light to break the shot-noise limit in particle tracking sensitivity~\cite{Taylor2012_squeezing}. In principle, this optical lock-in particle tracking scheme offers near immunity to low frequency laser noise and electronic noise, which could make it a highly practical method for a wide range of experiments. In the initial demonstration, however, the noise suppression attained through use of the optical lock-in tracking was not characterized~\cite{Taylor2012_squeezing}. Furthermore, the experimental setup in that demonstration was more complicated than necessary for classical particle tracking and incompatible with short working distance objectives. Here we demonstrate optical lock-in particle tracking with a simple optical setup which can be applied with any objectives. 
 It is shown that lock-in based particle tracking has superior sensitivity to continuous measurement at all frequencies where the continuous measurement is limited by technical noise, and achieves equivalent sensitivity where the dominant noise source is fundamental shot-noise. The reduction in laser noise and electronic noise yields on average 20~dB of noise suppression in the frequency range 10--5000~Hz, where  low frequency laser noise and electronic noise is significant, and 35~dB of noise suppression in the range 550--710~kHz where the laser crystal relaxation oscillations introduce a large noise feature.


\section{Basic concept}

\begin{figure}
 \begin{center}
   \includegraphics[width=8cm]{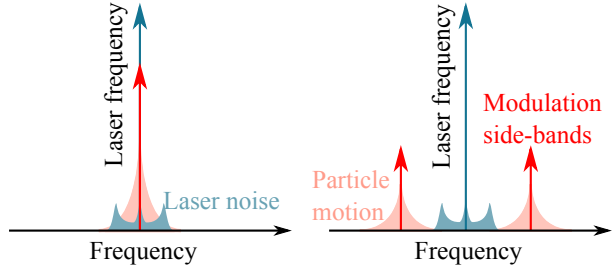}
   \caption{ A probe optical field is modulated by its interaction with the particle (red). In order to measure this, it is mixed with another bright local oscillator field (dark blue). However, the local oscillator also has some low-frequency noise present. If the probe field frequency matches the local oscillator frequency, as shown on the left, then the low-frequency noise competes with the low-frequency particle motion signal. However, if the probe is in amplitude modulated side-bands, as shown on the right, then the low-frequency  particle motion can be isolated from the low-frequency noise, thereby improving the measurement sensitivity.
}
 \label{Concept}  
 \end{center}
\end{figure}

The lock-in based particle tracking measurement demonstrated here is qualitatively similar to a continuous position tracking experiment. In optical tweezers based measurements, scattering particles are illuminated and the spatial distribution of the resulting scattered field is measured to infer particle position~\cite{Taylor2012QNL,Gittes1998}. Any modulation on the incident illumination is carried onto the scattered field, shifting some of the optical power from the laser carrier frequency into modulation side-bands. Once the scattered field is measured, the optical modulation translates into a modulation on the electrical signal, with the particle position information centered about the modulation frequency (see Fig.~\ref{Concept}). The particle position can be  recovered by demodulating this signal.

 We may ask how the expected sensitivity of such a measurement compares to a usual continuous measurement. When optical fields are measured, the resulting photocurrent at time $t$ is given by
 \begin{equation}
 I(t) = G \int U(X,Y) |E(t)|^2 dX\, dY + N_E(t)
 \end{equation}
 where $N_E$ is the electronic noise, $G$ is the detector gain, $E$ is the total electric field at the detector at the coordinates $X$ and $Y$,  and $U(X,Y)$ represents the spatial gain of the detector; for instance, if the photocurrent from two halves of a split detector are subtracted from one another, this is represented as $U(X,Y) = {\rm sign}(X)$, while a bulk detector has $U(X,Y)=1$. Here we assume that the fields present are a scattered field $E_{s}$ which depends on particle position, and a local oscillator $E_{\rm LO}$ with which the scattered field interferes, such that $E=E_LO + E_s$. In most optical tweezers experiments, the local oscillator is simply given by the component of the trapping field which has not scattered from the particle. For lock-in experiments, the fields are separated to allow the particle illumination to be modulated independently of the local oscillator. 
 The scattered field is assumed to be much smaller than the local oscillator ($|E_{s}| \ll |E_{\rm LO}| $) as is typically the case, such that the measured photocurrent is given by
%
%
 \begin{equation}
 I(t) = G \int  U(X,Y)|E_{\rm LO}(t)|^2  + 2 U(X,Y) {\rm Re} \{ E_{\rm LO}(t) E_{s}^*(t) \} dX\, dY +N_E(t). \label{I_def}
 \end{equation}
The explicit time dependence of the scattered field may be separated from the spatial mode shape as $E_{s} =( A_s(t) + \xi(t)) \psi_s(X,Y)$, where $A_s(t)$ and $\xi(t)$ are respectively the real expectation value of the field amplitude from which the particle tracking signal originates, and its fluctuations which contribute noise such as shot noise. $\psi_s(X,Y)$ is the complex spatial modeshape of the scattered field; this does not have explicit time dependence, but is modified as the particle moves. It is this spatial modification which is ultimately monitored to retrieve a particle tracking signal.  
%
 To find the dependence of the scattered field on a small particle displacement $x(t)$, it can be expanded to first order as 
 \begin{equation}
E_{s} = E_{s}|_{x=0} + x(t) \frac{d E_{s}}{d x}\big|_{x=0} = ( A_s(t) + \xi(t)) \psi_s(X,Y)|_{x=0} + x(t)( A_s(t) + \xi(t)) \frac{d \psi_s(X,Y)}{d x}\big|_{x=0}.
 \end{equation}
Substituting this expression into Eq.~\ref{I_def}, the component of the photocurrent which gives a linear particle tracking signal can be seen to be 
 \begin{eqnarray}
 I_{\rm sig} &=& 2 G x(t)  A_s(t) \int  U(X,Y) {\rm Re} \{ E_{\rm LO}  \frac{d \psi_s^*(X,Y)}{d x}\big|_{x=0} \} dX\, dY ,\label{I_sig}\\
 &=& g x(t) A_s(t)
 \end{eqnarray}
where for brevity we define a gain $g=2 G\int  U(X,Y){\rm Re} \{E_{\rm LO} \frac{d \psi_s^*(X,Y)}{d x}\big|_{x=0} \} dX\, dY$. The position sensitivity is optimized when this gain is maximized, as occurs when both the phase and shape of the local oscillator field are optimized to perfectly interfere with the scattered field component $ \frac{d E_{s}}{dx} \big|_{x=0} $~\cite{Taylor2012_squeezing,Tay2009}. Substituting this into Eq.~\ref{I_def}, we can represent the measured photocurrent as
 \begin{equation}
 I(t) = N_{\rm opt}(t) + N_E(t)  + g A_s(t) x(t) ,
 \end{equation}
where all the terms in the integrand which did not contribute to the tracking signal are included as optical noise $N_{\rm opt}$. 
%
%
 %
%
For a continuous measurement, the expectation value of the scattered field amplitude $A_s(t)$ should be constant. Alternatively, we can perform lock-in measurement if we modulate the scattered field amplitude at frequency $\omega$ such that  $A_s(t) = \sqrt{2}\bar{A}_{s} {\rm cos}(\omega t)$, where the modulated amplitude has an RMS value of $\bar{A}_{s}$. Provided the modulation frequency is much faster than the mechanical motion, the position can then be extracted by demodulation;
 \begin{equation}
  I_{\rm lock-in}=\sqrt{2} I {\rm cos}(\omega t) =  \sqrt{2}\left( N_{\rm opt}(t) + N_E(t)\right) {\rm cos}(\omega t) +  g \bar{A}_{s} x + g \bar{A}_{s} x {\rm cos}(2 \omega t). \label{Sens_Eq}
 \end{equation}
 Thus, the effect of the lock-in is to shift the low frequency noise to high frequencies, and generate a second harmonic term proportional to $x$ while leaving the signal term unchanged. The second harmonic term and the low frequency noise can then be removed via a low-pass filter, such that only the noise originally near the modulation frequency enters the measurement. 
 Wherever low-frequency noise is dominant, lock-in measurement allows suppression of the noise floor. This does not influence white noise sources such as shot-noise, since these are equally present at low frequencies and around the modulation frequency.  Thus, the fundamental shot-noise limit on position sensitivity is not influenced by a choice between continuous or lock-in measurement. The two schemes have equivalent shot-noise limits to sensitivity when the lock-in scattered amplitude $\bar{A}_{s}$ matches the amplitude ${A}_{s}$ of a the continuous measurment, or equivalently, when the same number of scattered photons in modulation side-bands are collected for the lock-in measurement as are collected for a continuous measurement.


\section{Experiment}

\begin{figure}
 \begin{center} 
   \includegraphics[width=10cm]{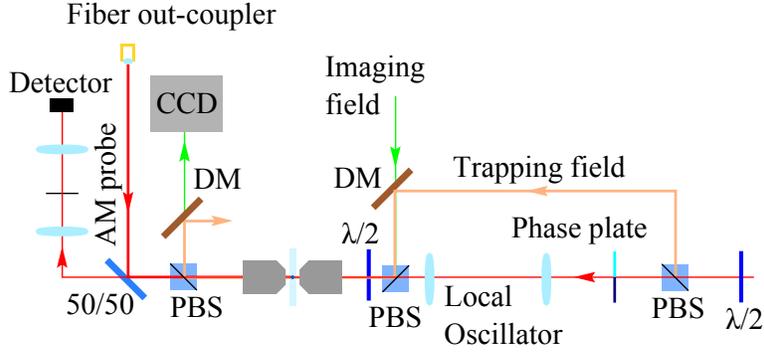}
   \caption{ Layout of the optical lock-in tracking scheme used here. PBS: polarizing beamsplitter, DM: dichroic mirror. The local oscillator is shaped with a phase plate which imparts a $\pi$ phase shift to one half of the spatial profile. Particles are trapped by the counter-propagating probe and trap fields. The trap field is isolated from the detection, and if it is not required for stable trapping, can be removed altogether. The probe field scatters from trapped particles, and the particle motion tracked via the interference between this scattered light and the local oscillator. The probe is amplitude modulated at 2~MHz in a fiber Mach-Zehnder modulator. A separate green field is used to image the particles in the trap.
}
 \label{Layout}  
 \end{center}
\end{figure}



\begin{figure}
 \begin{center}
   \includegraphics[width=12cm]{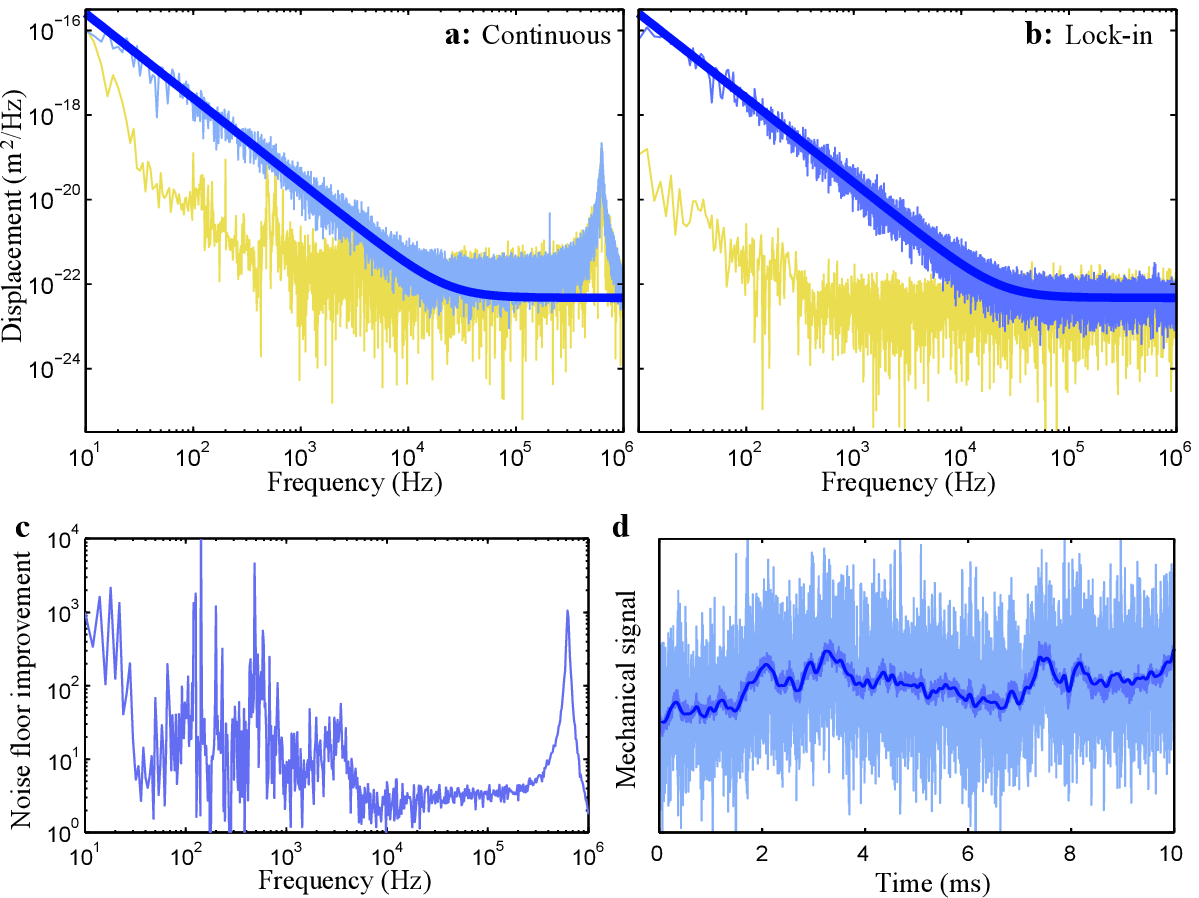}
   \caption{ Particle tracking spectra are shown from simultaneous continuous ({\bf a}) and lock-in ({\bf b}) measurements.  The light yellow trace shows the noise floor present in the absence of a trapped particle which corresponds to the measurement imprecision, and the blue shows the measured signal with a 1~$\mu$m polystyrene bead held in the trap. The thick darker blue line fits the bead motion and the flat shot-noise floor. This matches the lock-in data well since it is shot-noise limited from 500~Hz, but does not follow the continuous spectra as this was limited by low frequency laser noise until 1~MHz. This noise includes a very prominent spectral peak around 630~kHz from the laser diode relaxation oscillations. Because the fitted floor corresponds to the shot-noise level, it drops below the measured data between 10~kHz and 1~MHz. The trap was very weak, as we used 0.4 NA objectives with a total of 30~mW trapping field. As such, the corner frequency is slightly below 10~Hz and not visible in the displayed data.  By excluding low frequency noise, the lock-in measurement yields a measurement precision which is improved by the factor shown in {\bf c}. Subplot {\bf d} shows the continuous (light) and lock-in (medium) time-traces after a low-pass filter at 1~MHz, revealing the clear noise improvement from lock-in measurement, and also shows the continuous data with a low-pass filter at 10~kHz (dark), which closely follows the higher bandwidth lock-in results.
}
 \label{Plot}  
 \end{center}
\end{figure}

 The optical lock-in particle tracking experiment shown in Fig.~\ref{Layout} was built, and  the sensitivity attainable with continuous and lock-in measurements characterized.  A particle is trapped in water between two objectives with 0.4 numerical aperture (NA) by 1064~nm light produced by a low noise~\cite{Kwee2008} Innolight Prometheus Nd:YAG laser. Due to the low NA objectives used, trapping is not possible with a single beam. Two orthogonally polarized counter-propagating fields are used instead to confine particles, with only one of these contributing to the measurement. It is important to note that although a dual beam optical trap is used here, lock-in particle tracking is fully compatible with single-beam traps.  One of the trapping fields (referred to as the probe) is amplitude modulated at 2~MHz, which is sufficiently high that the resulting modulation of the trap strength does not measurably disturb the particle motion.  The back-scatter from this modulated probe field is combined with a local oscillator field which also propagates through the trap. The local oscillator is shaped with a phase plate so that particle motion modulates the spatial overlap between the local oscillator and scattered field. Provided the phase between the local oscillator and scattered field is correctly chosen, the influence of the particle motion on the interference between these fields directly maps the position onto the transmitted light intensity~\cite{Taylor2012_squeezing}, which is then measured on a New Focus 1811 bulk detector. Demodulation of the resulting signal at the amplitude modulation frequency allows both tracking of a scattering particle and also monitoring of the relative optical phases. The scattered light includes a large stationary term ($E_{s}|_{x=0}$ in Eq.~\ref{I_def}), and the phase between this and the local oscillator can be determined from the amplitude of the measured modulation. This measured phase was processed with a PID controller and locked by feedback to a piezo-mounted mirror in the path of the probe field. 
 This approach extends the scheme first developed for optical lock-in particle tracking~\cite{Taylor2012_squeezing} to a more typical optical trapping setup. There the scattered field was produced from side-illumination whereas here the modulated probe acts as a trapping field. Side-illumination is only possible if there is room for a free-space probe field to reach the trap center, which requires use of long working distance objectives (in Ref.~\cite{Taylor2012_squeezing}, 6~mm) which are not typically used for optical trapping.  Furthermore, the probe can only be weakly focused with side-illumination. Focusing the probe field through the objective increases its intensity by approximately $10^3$, which results in a corresponding increase in the scattered field amplitude and an improvement in the position sensitivity as described in Eq.~\ref{Sens_Eq}.  

 The amplitude modulation on the probe is chosen to leave approximately equal power in the central laser frequency and the first modulation side-band. This allows the continuous and pulsed measurements to occur simultaneously with a single detector, and with equivalent recording conditions. Some non-linearity in the modulator resulted in a number of higher harmonics being generated, which were suppressed in the data acquisition with analog electronic filters. 

 Using this setup, the Brownian motion of a 1~$\mu$m polystyrene bead was simultaneously measured both continuously and from side-bands around the 2~MHz modulation, with spectra shown in Fig.~\ref{Plot}{\bf a} and {\bf b} respectively. The background noise was characterized by performing equivalent measurements in the absence of a trapped bead. As expected, the lock-in measurement is very similar to the continuous measurement, but with a reduction in the included electronic and laser noise. The reduction in included noise (shown in Fig.~\ref{Plot}{\bf c}) causes the measurement imprecision to improve markedly at the frequencies where laser and electronic noise are dominant. Between 10 and 5000~Hz, the imprecision is improved by an average of 20~dB, with even greater suppression of 35~dB in the frequency range 550--710~kHz where the laser crystal relaxation oscillations produce a prominent laser noise peak centred at 630~kHz. A comparison of the two measurements in the time domain also reveals both the clear suppression of noise on the lock-in trace and the otherwise close agreement between the measured displacements (Fig.~\ref{Plot}{\bf d}). These results verify that the lock-in measurement is equivalent to a continuous measurement, except that it evades low frequency technical noise.


  With the optical layout used here, particle motion was tracked in a self-homodyne measurement on a single bulk detector rather than a quadrant photodiode. This is not required for lock-in particle tracking, which should work with any detection apparatus. However, it can be very advantageous; quadrant detectors are avoided in some high-speed experiments because they typically have low bandwidth~\cite{Chavez2008,Li2011}. Furthermore, the quantum limit on sensitivity is accessible only with perfect interference between the local oscillator and scattered fields, which requires the local oscillator to be spatially engineered, as it is in a homodyne measurement such as this~\cite{Taylor2012QNL,Tay2009}. 
 A difficulty with the layout used here was that some of the probe field reflected from the sample chamber into the detector. This back-reflection was of a greater intensity than the back-scatter from the particle, and phase shifts between the local oscillator and scattered fields generated a measured signal, but this was primarily below 5~Hz. In that low frequency range, our lock-in measurement performed worse than the continuous measurement, although eliminating the back-reflection with anti-reflection coatings would resolve this. Also, it should be noted that evasion of laser noise and electronic noise can only improve sensitivity to  the particle position relative to the optical fields. As with all other particle tracking experiments, this measurement remains sensitive to mirror drifts or air currents outside the trap which cause the trap center to drift,  and  conventional methods are needed to stabilize these noise sources.


We have demonstrated that a lock-in measurement scheme provides a simple and robust technique to reduce technical noise in an optical tweezers setup. This can yield a substantial improvement in sensitivity which could  be practical for many optical tweezers applications.

\section*{Acknowledgments}
 This work was supported by the Australian Research Council Discovery Project Contract No. DP0985078.

\end{document}